\documentclass[12pt,a4paper]{article}
\usepackage{lineno,hyperref}
\usepackage[latin5]{inputenc}
\usepackage{amsmath}
\usepackage{graphicx}
\usepackage{subfigure}
\usepackage{float}
\usepackage{mathtools}
\usepackage{color}
\usepackage{cite}
\usepackage{mathrsfs}
\usepackage[]{geometry}
\usepackage[singlelinecheck=false]{caption} 
\usepackage{graphicx}
\usepackage{cite}
\usepackage{authblk}
\usepackage{float}
\usepackage{color}
\usepackage{graphicx}
\usepackage{epstopdf}
\usepackage{amscd,amsmath,amssymb,amsthm,amsfonts,epsfig,graphics}
\DeclareMathOperator{\sech}{sech}
\DeclareMathOperator{\csch}{csch}
\newtheorem{theorem}{Theorem}
\usepackage{geometry}
\title{Traveling Wave Solutions to Conformable Time Fractional RLW-class equations}
\author{Gokhan Koyunlu \thanks{gokhankoyunlu@gmail.com}\\ 
Nile Univeristy of Nigeria,\\ 
Department of Computer Engineering ,Abuja , Nigeria.}

\begin{document}
\maketitle
\begin{abstract} 
The traveling wave solutions to some nonlinear conformable time fractional partial differential equations in RLW-class are set up by using $\sech$ and $\csch$ ansatzs. The conformable time fractional forms of the equal-width (EW), regularized long wave (RLW) and symmetric regularized long wave (sRLW) equations are considered in the study. By the assist of the simple traveling wave transformation, the equations are converted to some ordinary differential equations. Then, assuming these equations have solutions of forms of powers of $\sech$ and $\csch$ functions lead to determine the powers of the solutions if exist. The determination of the relation among the other parameters in the solutions follows the previous process. Finally, the solutions are expressed in some explicit forms. 
\end{abstract}
\textbf{\textit{AMS2010:}} 5C07;35R11;35Q53.\\

\section{Introduction}
The Regularized Long Wave (RLW) class equations are widely used models to comprehend many wave phenomena covering propagation, collusion or generation of waves. 

Probably the most known property of the RLW equation is to describe the growth of an undular bore from a long wave \cite{peregrine1}. The details, existence, uniqueness and stability, of some particular solutions of the RLW equation are deeply examined by Benjamin et al.\cite{benjamin1}. Wazwaz and Triki \cite{wazwaz1} generates some soliton solutions to a particular RLW equation with time dependent coefficients and dumping term.

Morrison et al. \cite{morrison1} state that it is not possible to integrate the RLW equation by the inverse scattering and the interaction of solitary waves is inelastic. They suggest the Lagrangian density for the RLW equation. The equal width (EW) equation was also described in the same study with some properties by using a simple transformation between the RLW and the EW equations. A large classification of single traveling wave type solutions to the generalized EW equation was given in Fan's study \cite{fan1}. Fan implemented the complete discrimination system for polynomial to the EW equation in the generalized form.

Seyler and Fenstermacher\cite{seyler1} derived the symmetric form of the RLW equation by implementing the Fourier transform to the cold-electron fluid equations.

Due to the difficulties to classify the non linear partial differential equations, a general solution technique for all nonlinear PDEs have not been developed yet. Instead, the integrability of the equations are investigated one by one. Even though most of them are reduced to some ordinary differential equations by using some particular transforms, it is not easy to integrate the resultant ODEs in many cases. Thus, the exact solution techniques have been derived to determine some solutions of the nonlinear PDEs. In many methods, a predicted solution with unknown parameters is substituted into the equation and the relation among the parameters are determined, if exists. The predicted solutions can be in series, rational or simple forms. In the recent studies, various techniques have been derived to solve different nonlinear partial differential equations covering the fractional ones\cite{zayed1,roshid1,kaplan1,eslami1,korkmaz1,ganji1,naher1,guner2,korkmaz2,guner3}. 

In the present study, the $\sech^B$ and $\csch^B$ type solutions to the time fractional forms of the RLW

\begin{equation}
D_{t}^{\alpha}u+pu+quu_x+ru_{xxt}=0 \label{cfrlw}
\end{equation}
the EW
\begin{equation}
D_{t}^{\alpha}u+puu_x+qu_{xxt}=0  \label{cfew}
\end{equation}
and the sRLW
\begin{equation}
D_{tt}^{2\alpha}u+pu_{xx}+quD_{t}^{\alpha}{u_x}+qu_xD_{t}^{\alpha}u+rD_{tt}^{2\alpha}{u_{xx}}=0	  \label{cfsrlw}
\end{equation}
equations are investigated. The derivative operator $D_{t}^{\alpha}$ denotes $\alpha$.th order conformable derivative with respect to the variable $t$ where $\alpha \in (0,1]$ as the subscripts involving $x$ denotes the classical derivative and $p$, $q$, and $r$ nonzero real parameters.

Before starting to construct the solutions, definition and some fundamental properties of the conformable derivative are given in the following section.

\section{Conformable Fractional Derivative}
Consider a function $u=u(t)$ defined in the positive side of the real line, $t>0$. The conformable fractional derivative of $u$ is defined as 
\begin{equation}
D_{t}^{\alpha}(u(t))=\lim\limits_{h\rightarrow 0}{\dfrac{u(t+h t^{1-\alpha})-u(t)}{h}}
\end{equation}
where $\alpha \in (0,1]$ \cite{khalil1}. The conformable fractional derivative satisfies the following properties for $u=u(t)$ and $v=v(t)$ are sufficiently $\alpha$-differentiable for all $t>0$:

\begin{itemize}
\item $D_{t}^{\alpha}(c_1u+c_2v)=c_1D_{t}^{\alpha}(u)+c_2D_{t}^{\alpha}(v)$ 
\item $D_{t}^{\alpha}(t^k)=kt^{k-\alpha}, \forall k \in \mathbb{R}$
\item $D_{t}^{\alpha}(\lambda)=0$, $ \lambda$ constant
\item $D_{t}^{\alpha}(uv)=uD_{t}^{\alpha}(v)+vD_{t}^{\alpha}(u)$
\item $D_{t}^{\alpha}(\frac{u}{v})=\dfrac{vD_{t}^{\alpha}(u)-uD_{t}^{\alpha}(v)}{v^2}$
\item $D_{t}^{\alpha}(u)(t)=t^{1-\alpha}\frac{du}{dt}$
\end{itemize}
for $\forall c_1,c_2 \in \mathbb{R}$\cite{atangana1}. 
\begin{theorem}
Let $u=u(t)$ be an $\alpha$-conformable differentiable function, and $v=v(t)$ is also differentiable function defined in the range of $u$. Then,
\begin{equation}
D_{t}^{\alpha}(u\circ v)(t)=t^{1-\alpha}v^{\prime}(t)u^{\prime}(v(t))
\end{equation}
where $^{\prime}$ denotes derivative\cite{abdeljawad1}.
\end{theorem}

\section{The Procedure of Solution}
A general fractional partial differential equation
\begin{equation}
P_1(u,D_{t}^{\alpha}u,u_x,D_{t}^{\alpha}u_x,u_{xx},\ldots )=0 \label{fpde}
\end{equation}
is reduced to 
\begin{equation}
	P_2(U,U',U'',\ldots )=0 \label{oode}
\end{equation}
where $'$ denotes the ordinary derivative
by using the traveling wave transformation
\begin{equation}
	u(x,t)=U(\xi), \xi = ax-\frac{\nu}{\alpha}t^{\alpha} \label{wt}
\end{equation}
In the transformation (\ref{wt}), $a$ and $\nu$ are nonzero constants. Now assume that (\ref{oode}) has a solution of the form
\begin{equation}
	U(\xi)=Af(\xi) \label{solode}
\end{equation}
where $f(\xi)=\sech^B(\xi)$ or $f(\xi)=\csch^B(\xi)$, $A\neq 0$ and $B\in \mathbb{Z}^{+}$ are nonzero constants. Substituting the chosen solution into (\ref{oode}) and rearranging the resultant equation for powers of $\sech$ or $\csch$ function leads a simple way to determine $B$. If it is managed to find a positive integer $B$, the next step can be started. Otherwise, the procedure fails. After determination of $B$, the solution (\ref{solode}) is substituted into (\ref{oode}) with the determined value of $B$. Using the polynomial equation for $\sech$ or $\csch$ functions yields a system of algebraic equations. The system is solved to find the relation among the parameters $A$, $\nu$ and $a$.

\section{Traveling Wave Solutions to Conformable Time Fractional RLW Equation}
The traveling wave transformation (\ref{wt}) reduces the conformal time fractional form of the RLW equation (\ref{cfrlw}) to
\begin{equation}
	-\nu\,{\frac {\rm d}{{\rm d}\xi}}u \left( \xi \right) +pa{\frac 
{\rm d}{{\rm d}\xi}}u \left( \xi \right) +qau \left( \xi \right) {
\frac {\rm d}{{\rm d}\xi}}u \left( \xi \right) -\nu\,r{a}^{2}{\frac {
{\rm d}^{3}}{{\rm d}{\xi}^{3}}}u \left( \xi \right) =0
\end{equation}
Integrating once this equation gives
\begin{equation}
	-\nu\,u\left( \xi \right)+pau\left( \xi \right)+\frac{1}{2}\,qau\left( \xi \right)-\nu\,r{a}^{2}{\frac {{\rm d}^{2}}{{\rm d}{\xi}^{2
}}}u \left( \xi \right) =K \label{rlwode}
\end{equation}
where $K$ is the constant of integration. The procedure for both $\sech^B$ and $\csch^B$-type solutions is the same to this point.
\subsection{sech-type solutions}
Substituting the predicted solution $A\sech^B\left( \xi \right)$ to (\ref{rlwode}) gives
\begin{equation}
	\frac{1}{2}\,{A}^{2}aq \sech^{2\,B}\xi+
 \left( -\nu\,r{a}^{2}A{B}^{2}+paA-\nu\,A \right) \sech^{B}\xi+ \left( \nu\,r{a}^{2}A{B}^{2}+AB{a}^{2}
\nu\,r \right)  \sech^{B+2}\xi  =K
\end{equation}
Equating $2B$ to $B+2$ gives $B=2$. Thus, the solution should be in the form $A\sech^2\xi$. Substituting $A\sech^2\xi$ into (\ref{rlwode}) leads to
\begin{equation}
	\left( \frac{1}{2}\,qa{A}^{2}+6\,\nu\,r{a}^{2}A \right)  \sech^{4}\xi+ \left( -4\,\nu\,r{a}^{2}A+paA-\nu\,A
 \right)  \sech^{2}\xi-K=0
\end{equation}
Since we look for a nonzero solution, the coefficients of the powers of $\sech\xi$ should be zero with $K=0$. Solving the algebraic equation system
\begin{equation}
	\begin{aligned}
	\frac{1}{2}\,qa{A}^{2}+6\,\nu\,r{a}^{2}A &=0\\
	-4\,\nu\,r{a}^{2}A+paA-\nu\,A&=0
	\end{aligned}
\end{equation}
for $A$ and $a$ gives
\begin{equation}
	\begin{aligned}
	A&=\pm \frac{3}{2}\,{\frac {p+\sqrt {-16\,{\nu}^{2}r+{p}^{2}}}{q}}\\
	a&= \mp \frac{1}{8}\,{\frac {p+\sqrt {-16\,{\nu}^{2}r+{p}^{2}}}{\nu\,r}}
	\end{aligned}
\end{equation}
where $q\neq 0$ and $r\neq 0$. One should also state that the existence of real solutions depends on one more additional condition on $r$ as $r<0$. Thus, the solutions to (\ref{rlwode}) are developed as
\begin{equation}
	\begin{aligned}
	U_1(\xi)&=\frac{3}{2}\,{\frac {p+\sqrt {-16\,{\nu}^{2}r+{p}^{2}}}{q}} \sech^2{\xi}\\
	U_2(\xi)&=-\frac{3}{2}\,{\frac {p+\sqrt {-16\,{\nu}^{2}r+{p}^{2}}}{q}} \sech^2{\xi}
	\end{aligned}
\end{equation}
These solutions are written as
\begin{equation}
	\begin{aligned}
	u_1(x,t)&=\frac{3}{2}\,{\frac {p+\sqrt {-16\,{\nu}^{2}r+{p}^{2}}}{q}} \sech^2{\left(-\frac{1}{8}\,{\frac {p+\sqrt {-16\,{\nu}^{2}r+{p}^{2}}}{\nu\,r}}x-\nu\frac{t^{\alpha}}{\alpha}\right)}\\
	u_2(x,t)&=-\frac{3}{2}\,{\frac {p+\sqrt {-16\,{\nu}^{2}r+{p}^{2}}}{q}} \sech^2{\left(\frac{1}{8}\,{\frac {p+\sqrt {-16\,{\nu}^{2}r+{p}^{2}}}{\nu\,r}}x-\nu\frac{t^{\alpha}}{\alpha}\right)}
	\end{aligned}
\end{equation}
for arbitrarily chosen $\nu \neq 0$, $r\neq 0$ and $q\neq 0$.
\subsection{csch-type solutions}
Let $A\csch^B{\xi}$ be a solution to (\ref{rlwode}). Then, substituting it into (\ref{rlwode}) gives
\begin{equation}
	\frac{1}{2}\,{A}^{2}aq \csch^{2\,B}\xi+
 \left( -\nu\,r{a}^{2}A{B}^{2}+paA-\nu\,A \right)  \csch^{B}\xi+ \left( -\nu\,r{a}^{2}A{B}^{2}-AB{a}^{2}
\nu\,r \right)  \csch^{B+2}\xi=K
\end{equation}
Choosing $2B=B+2$ results in $B=2$. Thus, the predicted solution takes the form $A\csch^2(\xi)$. Substituting this solution into (\ref{rlwode}) leads
\begin{equation}
	 \left( \frac{1}{2}\,qa{A}^{2}-6\,\nu\,r{a}^{2}A \right) \csch^{4}\xi+ \left( -4\,\nu\,r{a}^{2}A+paA-\nu\,A
 \right) \csch^{2}\xi-K=0
\end{equation}
This equation is valid only for the coefficients of the powers of $\csch$ and $K$ both are zero under the nonzero solution assumption of (\ref{rlwode}). It is clear that the relation between the parameters in both the traveling wave transformation (\ref{wt}) and the predicted solution can be determined by solving the algebraic system
\begin{equation}
	\begin{aligned}
	 \frac{1}{2}\,qa{A}^{2}-6\,\nu\,r{a}^{2}A&=0 \\
	-4\,\nu\,r{a}^{2}A+paA-\nu\,A&=0
	\end{aligned}
\end{equation}
Solving this system for $A$ and $a$ gives
 \begin{equation}
	\begin{aligned}
	 A&=24\,{\frac {\nu\,r}{ \left( p+\sqrt {-16\,{\nu}^{2}r+{p}^{2}}
 \right) q} \left( \frac{1}{8}\,{\frac {p \left( p+\sqrt {-16\,{\nu}^{2}r+{p
}^{2}} \right) }{\nu\,r}}-\nu \right) } \\
	a&=\frac{1}{8}\,{\frac {p+\sqrt {-16\,{\nu}^{2}r+{p}^{2}}}{\nu\,r}}
	\end{aligned}
\end{equation}
and 
 \begin{equation}
	\begin{aligned}
	 A&=-24\,{\frac {\nu\,r}{ \left( -p+\sqrt {-16\,{\nu}^{2}r+{p}^{2}}
 \right) q} \left( -\frac{1}{8}\,{\frac {p \left( -p+\sqrt {-16\,{\nu}^{2}r+{p
}^{2}} \right) }{\nu\,r}}-\nu \right) } \\
	a&=-\frac{1}{8}\,{\frac {-p+\sqrt {-16\,{\nu}^{2}r+{p}^{2}}}{\nu\,r}}
	\end{aligned}
\end{equation}
for arbitrarily chosen nonzero $\nu$. Thus, the solution to (\ref{rlwode}) is constructed as
\begin{equation}
	\begin{aligned}
	U_3(\xi)&= 24\,{\frac {\nu\,r}{ \left( p+\sqrt {-16\,{\nu}^{2}r+{p}^{2}}
 \right) q} \left( \frac{1}{8}\,{\frac {p \left( p+\sqrt {-16\,{\nu}^{2}r+{p
}^{2}} \right) }{\nu\,r}}-\nu \right) }\csch^2(\xi) \\
U_4(\xi)&=-24\,{\frac {\nu\,r}{ \left( -p+\sqrt {-16\,{\nu}^{2}r+{p}^{2}}
 \right) q} \left( -\frac{1}{8}\,{\frac {p \left( -p+\sqrt {-16\,{\nu}^{2}r+{p
}^{2}} \right) }{\nu\,r}}-\nu \right) } \csch^2(\xi)
	\end{aligned}
\end{equation}
for $q\neq 0$, $r \neq 0$ and $\pm p+ \sqrt{16\,{\nu}^{2}r+{p}^{2}} \neq 0$. Returning the original variables $x$ and $t$ converts the solutions $U_3(\xi)$ and $U_4(\xi)$ to the solutions to (\ref{cfrlw}) as
{\scriptsize
\begin{equation}
	\begin{aligned}
	u_3(x,t)&= 24\,{\frac {\nu\,r}{ \left( p+\sqrt {-16\,{\nu}^{2}r+{p}^{2}}
 \right) q} \left( \frac{1}{8}\,{\frac {p \left( p+\sqrt {-16\,{\nu}^{2}r+{p
}^{2}} \right) }{\nu\,r}}-\nu \right) }\csch^2(\frac{1}{8}\,{\frac {p+\sqrt {-16\,{\nu}^{2}r+{p}^{2}}}{\nu\,r}}x-\nu\frac{t^{\alpha}}{\alpha}) \\
u_4(x,t)&=-24\,{\frac {\nu\,r}{ \left( -p+\sqrt {-16\,{\nu}^{2}r+{p}^{2}}
 \right) q} \left( -\frac{1}{8}\,{\frac {p \left( -p+\sqrt {-16\,{\nu}^{2}r+{p
}^{2}} \right) }{\nu\,r}}-\nu \right) } \csch^2(-\frac{1}{8}\,{\frac {-p+\sqrt {-16\,{\nu}^{2}r+{p}^{2}}}{\nu\,r}}x-\nu \frac{t^{\alpha}}{\alpha})
	\end{aligned}
\end{equation}}
\section{Traveling Wave Solutions to Conformable Time Fractional EW Equation}
The traveling wave transformation given in (\ref{wt}) reduces the conformable time fractional EW equation (\ref{cfew})to
\begin{equation}
	-\nu u(\xi)+\frac{1}{2}pau^2(\xi)-\nu qa^2\frac{d^2}{d\xi^2}(u(\xi))-K=0 \label{ewode}
\end{equation}
where $K$ is the integration constant.
\subsection{sech-type solutions}
When the predicted solution $A\sech^B(\xi)$ is substituted into (\ref{ewode}), it reduces
\begin{equation}
	\frac{1}{2}\,{A}^{2}ap\sech^{2\,B}\xi+
 \left( -\nu\,q{a}^{2}A{B}^{2}-\nu\,A \right)  \sech^{B}\xi+ \left( \nu\,q{a}^{2}A{B}^{2}+AB{a}^{2}
\nu\,q \right) \sech^{B+2}\xi-K=0
\end{equation}
Equating $2B$ to $B+2$ gives the suitable $B$ as $2$. Thus, the predicted solution takes the form $A\sech^2{\xi}$. Substitution of this solution into (\ref{ewode}) leads
\begin{equation}
	\left( \frac{1}{2}\,pa{A}^{2}+6\,\nu\,q{a}^{2}A \right)  \sech^{4}\xi+ \left( -4\,\nu\,q{a}^{2}A-\nu\,A
 \right)  \sech^{2}\xi-K=0
\end{equation}
It is clear that the solution of this equation for nonzero $\sech(\xi)$ function requires 
\begin{equation}
	\begin{aligned}
	 \frac{1}{2}\,pa{A}^{2}+6\,\nu\,q{a}^{2}A&=0\\
	-4\,\nu\,q{a}^{2}A-\nu\,A&=0\\
	K&=0
	\end{aligned}
\end{equation}
The solution of this algebraic system yields the relations among the parameters in the traveling wave transformation and the predicted solution as
\begin{equation}
	\begin{aligned}
	 A&=12\,{\frac {\nu\,q}{p}\sqrt {-\frac{1}{4q}}}\\
	a&=-\sqrt {-\frac{1}{4q}}
	\end{aligned}
\end{equation}
and
\begin{equation}
	\begin{aligned}
	 A&=-12\,{\frac {\nu\,q}{p}\sqrt {-\frac{1}{4q}}}\\
	a&=\sqrt {-\frac{1}{4q}}
	\end{aligned}
\end{equation}
where $p$ and $q$ are both nonzero. Following these values, the solutions to (\ref{ewode}) can be written as
\begin{equation}
	 U_{5,6}(\xi)=\pm 12\,{\frac {\nu\,q}{p}\sqrt {-\frac{1}{4q}}} \sech^2{\xi}
\end{equation}
Returning the original variables $x$, $t$ and $u(x,t)$ converts these solutions to
\begin{equation}
	\begin{aligned}
	 u_5(x,t)&=12\,{\frac {\nu\,q}{p}\sqrt {-\frac{1}{4q}}}\sech^2{\left(-\sqrt {-\frac{1}{4q}}x-\nu \frac{t^{\alpha}}{\alpha}\right)}\\
 u_6(x,t)&=-12\,{\frac {\nu\,q}{p}\sqrt {-\frac{1}{4q}}}\sech^2{\left(\sqrt {-\frac{1}{4q}}x-\nu \frac{t^{\alpha}}{\alpha}\right)}
	\end{aligned}
\end{equation}
\subsection{csch-type solutions}
Let $A\csch^B(\xi)$ be a nonzero solution to (\ref{ewode}). Then, substitution of this solution into (\ref{ewode}) gives
\begin{equation}
	1/2\,{A}^{2}ap \csch^{2\,B}\xi+
 \left( -qA{a}^{2}\nu\,{B}^{2}-\nu\,A \right)  \csch^{B}\xi+ \left( -qA{a}^{2}\nu\,{B}^{2}-qA{a}^{2}
\nu\,B \right)  \csch^{B+2}\xi-K=0
\end{equation}
Equating $2B$ to $B+2$ gives $B=2$. Then, the predicted solution can be written in the form $\csch^2\xi$. Putting this solution into (\ref{ewode}) leads
\begin{equation}
	 \left(\frac{1}{2}\,p{A}^{2}a-6\,qA{a}^{2}\nu \right) \csch^{4}\xi+ \left( -4\,qA{a}^{2}\nu-\nu\,A \right) 
 \csch^{2}\xi-K=0
\end{equation}
The solution of the algebraic system setup from the coefficients of the powers of $\csch$ and $K$ gives
\begin{equation}
	\begin{aligned}
		A&=12\,\frac {\nu\,q}{p}\sqrt {-\frac{1}{4q}} \\
		a&=\sqrt {-\frac{1}{4q}} \\
		K&=0
	\end{aligned}
\end{equation}
and
\begin{equation}
	\begin{aligned}
		A&=-12\,\frac {\nu\,q}{p}\sqrt {-\frac{1}{4q}} \\
		a&=-\sqrt {-\frac{1}{4q}}\\
		K&=0
	\end{aligned}
\end{equation}
with constraints $p \neq 0$ and $q \neq 0$. Thus, the solutions to (\ref{ewode}) can be written as
\begin{equation}
		U_{7,8}(\xi)=\pm 12\,{\frac {\nu\,q}{p}\sqrt {-\frac{1}{4q}}} \csch^2\xi
\end{equation}
Returning the original variables gives the solution to the conformable time fractional EW equation as
\begin{equation}
	\begin{aligned}
		u_{7}(x,t)&=12\,{\frac {\nu\,q}{p}\sqrt {-\frac{1}{4q}}} \csch^2\left(\sqrt {-\frac{1}{4q}}x-\nu \frac{t^{\alpha}}{\alpha} \right)\\
	u_8(x,t)&=-12\,{\frac {\nu\,q}{p}\sqrt {-\frac{1}{4q}}} \csch^2\left(-\sqrt {-\frac{1}{4q}}x-\nu \frac{t^{\alpha}}{\alpha} \right)
	\end{aligned}
\end{equation}
for arbitrarily chosen nonzero $\nu$.
\section{Traveling Wave Solutions to Conformable Time Fractional sRLW Equation}
The traveling wave transform (\ref{wt})  and integration of the resultant ordinary differential equation reduces the conformable fractional sRLW equation (\ref{cfsrlw}) to
\begin{equation}
	 \left( p{a}^{2}+{\nu}^{2} \right) {\frac {\rm d}{{\rm d}\xi}}u
 \left( \xi \right) -rqau{\frac {\rm d}{{\rm d}\xi}}u \left( \xi
 \right) +{\nu}^{2}{a}^{2}r{\frac {{\rm d}^{3}}{{\rm d}{\xi}^{3}}}u
 \left( \xi \right) =0 \label{srlwode}
\end{equation}
when the constant of integration is assumed zero.
\subsection{sech-type solution}
Assuming (\ref{srlwode}) has a solution of the form $A\sech^B\xi$ and substituting this solution into (\ref{srlwode}) results in
{\scriptsize
\begin{equation}
	{A}^{2}Baqr \sech^{2\,B}\xi+ \left( 
-A{B}^{3}{a}^{2}{\nu}^{2}r-AB{a}^{2}p-AB{\nu}^{2} \right) \sech^{B}\xi+ \left( A{B}^{3}{a}^{2}{\nu}^{
2}r+3\,A{B}^{2}{a}^{2}{\nu}^{2}r+2\,AB{a}^{2}{\nu}^{2}r \right) 
 \sech^{B+2}\xi=0
\end{equation}}
Equating $2B$ to $B+2$ gives $B=2$. Thus, the predicted solution reduces to $A\sech^2\xi$. Substituting this solution into (\ref{srlwode}) gives
\begin{equation}
	 \left( 24\,{\nu}^{2}{a}^{2}rA+2\,{A}^{2}aqr \right)  \sech^{4}\xi + \left( -8\,{\nu}^{2}{a}^{2}rA
-2\,pA{a}^{2}-2\,A{\nu}^{2} \right) \sech^{2}\xi=0
\end{equation}
Since $\sech\xi$ is nonzero, this equation is satisfied when
\begin{equation}
	\begin{aligned}
	24\,{\nu}^{2}{a}^{2}rA+2\,{A}^{2}aqr&=0\\
	-8\,{\nu}^{2}{a}^{2}rA
-2\,pA{a}^{2}-2\,A{\nu}^{2}&=0
	\end{aligned}
\end{equation}
Solving this algebraic system of equations for $A$ and $a$ gives
\begin{equation}
	\begin{aligned}
	A&=12\,{\frac {{\nu}^{3}}{q}\sqrt {- \frac{1}{4\,{\nu}^{2}r+p }
}}\\
	a&=-\sqrt {- \frac{1}{4\,{\nu}^{2}r+p }}\nu
	\end{aligned}
\end{equation}

and
\begin{equation}
	\begin{aligned}
	A&=-12\,{\frac {{\nu}^{3}}{q}\sqrt {- \frac{1}{4\,{\nu}^{2}r+p }
}}\\
	a&=\sqrt {- \frac{1}{4\,{\nu}^{2}r+p }}\nu
	\end{aligned}
\end{equation}
where $p\neq 0$ and $4\,{\nu}^{2}r+p \neq 0$. Thus, the solutions to (\ref{srlwode}) is constructed as

\begin{equation}
	U_{9,10}=\pm 12\,{\frac {{\nu}^{3}}{q}\sqrt {- \frac{1}{4\,{\nu}^{2}r+p }
}} \sech^2\xi
\end{equation}
Returning the original variables gives the solutions to (\ref{cfsrlw}) as
\begin{equation}
	\begin{aligned}
	u_9(x,t)&=-12\,{\frac {{\nu}^{3}}{q}\sqrt {- \frac{1}{4\,{\nu}^{2}r+p }}}\sech^2\left( \sqrt {- \frac{1}{4\,{\nu}^{2}r+p }}\nu x -\nu \frac{t^{\alpha}}{\alpha} \right)\\
u_{10}(x,t)&=12\,{\frac {{\nu}^{3}}{q}\sqrt {- \frac{1}{4\,{\nu}^{2}r+p }
}}\sech^2\left( -\sqrt {- \frac{1}{4\,{\nu}^{2}r+p }}\nu x -\nu \frac{t^{\alpha}}{\alpha} \right)
	\end{aligned}
\end{equation}

\subsection{csch-type solutions}
When the predicted solution $A\csch^B\xi$ is substituted into (\ref{srlwode}), the equation becomes
\begin{equation}
	{A}^{2}Baqr \csch^{2\,B}\xi + \left( -A{B}^{3}{a}^{2}{\nu}^{2}r-AB{a}^{2}p-AB{\nu}^{2} \right) \csch^{B}\xi + \left( -A{B}^{3}{a}^{2}{\nu}^
{2}r-3\,A{B}^{2}{a}^{2}{\nu}^{2}r-2\,AB{a}^{2}{\nu}^{2}r \right) 
 \csch^{B+2}\xi=0
\end{equation}
Equating $2B$ to $B+2$ results in $B=2$. Then, the predicted solution can be represented as $A\csch^2\xi$. Putting this form of the predicted solution into (\ref{srlwode}) results in
\begin{equation}
	 \left( -24\,A{a}^{2}{\nu}^{2}r+2\,{A}^{2}aqr \right) \csch^{4}\xi+ \left( -8\,A{a}^{2}{\nu}^{2}r
-2\,A{a}^{2}p-2\,A{\nu}^{2} \right) \csch^{2}\xi=0
\end{equation}
Similarly, the solution of the algebraic system of equations derived by equating the coefficients of the powers of $\csch$ to zero is determined as

\begin{equation}
	\begin{aligned}
	A&=12\,{\frac {{\nu}^{3}}{q}\sqrt {- \frac{1}{ 4\,{\nu}^{2}r+p}}}\\
a&=\sqrt {- \frac{1}{ 4\,{\nu}^{2}r+p}}\nu
	\end{aligned}
\end{equation}
and
\begin{equation}
	\begin{aligned}
	A&=-12\,{\frac {{\nu}^{3}}{q}\sqrt {- \frac{1}{ 4\,{\nu}^{2}r+p}}}\\
a&=-\sqrt {- \frac{1}{ 4\,{\nu}^{2}r+p}}\nu
	\end{aligned}
\end{equation}
Thus, the solutions to (\ref{srlwode}) can be represented as
\begin{equation}
	U_{11,12}(\xi)=\pm 12\,{\frac {{\nu}^{3}}{q}\sqrt {- \frac{1}{ 4\,{\nu}^{2}r+p}}} \csch^2\xi
\end{equation}
These solutions can be rewritten as
\begin{equation}
	\begin{aligned}
	u_{11}(x,t)=12\,{\frac {{\nu}^{3}}{q}\sqrt {- \frac{1}{ 4\,{\nu}^{2}r+p}}} \csch^2\left( \sqrt {- \frac{1}{ 4\,{\nu}^{2}r+p}}\nu x-\nu \frac{t^{\alpha}}{\alpha}\right)\\
	u_{12}(x,t)=-12\,{\frac {{\nu}^{3}}{q}\sqrt {- \frac{1}{ 4\,{\nu}^{2}r+p}}} \csch^2\left( -\sqrt {- \frac{1}{ 4\,{\nu}^{2}r+p}}\nu x-\nu \frac{t^{\alpha}}{\alpha}\right)
	\end{aligned}
\end{equation}
for arbitrarily chosen $\nu$.
\section{Conclusion}
$\sech^B$ and $\csch^B$-type solutions are determined for the conformable fractional RLW, EW and sRLW equations. The compatible traveling wave transformation defined in one space dimension reduces these equations to related ordinary differential equations. The determination of the power parameter is accomplished initially by substituting the predicted solution into the related ordinary differential equation. Once the power parameter is determined, the determination of the relations among the other parameters and constraints are done by algebraic method. The solutions of all three conformable time fractional partial differential equations are represented explicitly.

\end{document}